\documentclass[12pt]{article}
\usepackage{graphicx,floatflt,amssymb,epsf,rotate} 
\begin{document}
\thispagestyle{empty}
%\begin{flushright}
%\texttt{hep-ph/yymmnnn} \\
%\texttt{HRI-P-07-02-PQR}\\
%\texttt{CU-PHYSICS-xx/2007}\\
%\end{flushright}
\vskip 50pt
\parindent 0pt

\begin{center}
{\large {\bf SU(6), Triquark states, and the pentaquark}}

\vskip 10pt
\renewcommand{\thefootnote}{\fnsymbol{footnote}}

{\sf Swarup Kumar Majee\footnote{E-mail address:
swarup@mri.ernet.in}} 
and 
{\sf Amitava Raychaudhuri}

\vskip 8pt

{\em Harish-Chandra Research Institute, \\ Chhatnag Road, Jhunsi,
Allahabad 211 019, India\\and \\Department of Physics, University
of Calcutta,\\ 92 Acharya Prafulla Chandra Road, Kolkata 700 009,
India }

\vskip 60pt

{\bf ABSTRACT}

\end{center}

The purported observation of a state $\Theta^+$ with strangeness
S = +1 led to its quark model interpretation in terms of a
pentaquark combination involving a triquark-diquark structure --
the Karliner-Lipkin model.  In this work,  the proper colour-spin
symmetry properties for the $q q \bar{q}$ triquark are elucidated
by calculating the SU(6) unitary scalar factors and Racah
coefficients. Using these results, the colour-spin hyperfine
interactions, including flavour symmetry breaking therein, become
straight-forward to incorporate and the pentaquark masses are
readily obtained.    We examine the effect on the pentaquark mass
of (a) deviations from the flavour symmetric limit and (b)
different strengths of the doublet and triplet hyperfine
interactions. Reference values of these parameters  yield a
$\Theta^+$ mass prediction of 1601 MeV but it can comfortably
accommodate 1540 MeV for alternate choices. In the same framework, other
pentaquark states $\Xi$ (S=--2) and $\Theta^c $ (with charm
C=--1) are expected at 1783 MeV and 2757 MeV, respectively.

\vskip 5pt

\begin{center} 
\end{center}

\newpage

\renewcommand{\thesection}{\Roman{section}}
\setcounter{footnote}{0}
\renewcommand{\thefootnote}{\arabic{footnote}}

\section{Introduction}

Since long, baryon spectroscopy has been an arena to learn about
low-energy quantum chromodynamics. The purported observation of a
narrow baryon state of strangeness +1 at a mass around 1540 MeV,
$\Theta^+$, by several experiments \cite{theta} brought renewed
attention to this theatre. The evidence in support of this new
state is now of conflicting nature, loaded more in the direction
of non-observation \cite{nonob, rev}. Within the quark picture, the
positive strangeness ($\equiv \bar{s}$) of the $\Theta^+$ baryon
puts it in an exotic category and entails an interpretation in
terms of a minimum of four quarks and an antiquark -- a
pentaquark state ($udud\bar{s}$).

Soon after, three other states which also demand a pentaquark
classification were also observed.  These are the $\Xi^{--}
(dsds\bar{u})$ and $\Xi^{0} (dsus\bar{d})$ both at 1862 MeV
\cite{na49} and the $\Sigma^c (udud\bar{c})$ \cite{hera} with
mass 3099 MeV.

Though exotic states such as the pentaquark have a long history,
particular attention was drawn to a possible $\Theta^+$-like
state in the SU(3) version of the chiral soliton model
\cite{soliton}. Subsequently, the experimental results have
stimulated the exploration of many ideas, e.g., quark clusters,
colour hyperfine interactions,   Goldstone boson exchange, QCD
sum rules, lattice methods, etc., which have been reviewed in the
literature \cite{jm}.

For the $\Theta^+$, within the quark model framework, two models
\cite{kl,jw} have achieved special prominence. It is convenient to
discuss these using the language of SU(6) of colour-spin, SU(3)
of colour, and SU(2) of spin. Thus, for example, a quark
transforms as (6,3,2), where the three integers within the
parentheses identify the representations of the above SU(6),
SU(3), and SU(2), respectively. To avoid cluttering, the flavour
SU(3) structure is not explicitly shown. Our interest will
be on the triquark state which is an ingredient of the
Karliner-Lipkin model \cite{kl}.

An alternative possibility is the Jaffe--Wilczek (JW) model
\cite{jw}. Here the four quarks are assumed to form
two diquark clusters, each in the $(21,\bar{3},1)$
representation.  Of the four possible combinations
for a two-quark cluster -- (21,6,3), (15,6,1), $(15,\bar{3},3)$,
$(21,\bar{3},1)$ -- this is the one of the lowest energy. The two
diquark clusters and the remaining antiquark -- each one of which is
in colour $\bar{3}$ -- combine to form the colour singlet
pentaquark state $(qq)(qq)(\bar{q})$, {\em e.g.,} $\Theta^+
\equiv (ud)(ud)(\bar{s})$. A relative orbital angular momentum,
L=1, is assumed between the diquarks; this is in tune with the
observed narrow width of the state.  Another consequence is that
the pentaquark parity  
is predicted to be positive.  Note that the colour-spin symmetric
nature of the $(21,\bar{3},1)$ diquark requires it to be
antisymmetric, $\bar{3}$, in flavour to satisfy the generalized
Pauli principle.  The two diquarks (colour $\bar{3}$ bosons)
combine to form colour 3 to match up with the antiquark. This,
and L=1, requires the combination to be in a flavour symmetric
$\bar{6}$ state. The overall pentaquark flavour must be in
$\bar{6} \otimes \bar{3} = 8 + \overline{10}$.  The quantum
numbers of $\Theta^+$ can be accommodated only in the
$\overline{10}$.

In the Karliner-Lipkin (KL) model the quark clustering is
different. Here, it is postulated that there is one diquark cluster
with the same quantum numbers as in the JW model. The difference
is that the remaining two quarks and the antiquark are assumed to
form a triquark cluster $(qq\bar{q})$ with the quantum numbers
(6,3,2) which is in a flavour $\bar6$. The pentaquark state is
the colour singlet  $(qq)(qq\bar{q})$ combination. To explain the
narrowness of the observed states, a relative orbital angular
momentum, L=1, is postulated between the clusters so that the
parity of the state is predicted to be positive in this model as
well. The flavour structure of the states is the same as in the
JW model.

In this work, we set two goals. First, we take a detailed look at
the group theoretic properties of the triquark state.  We derive
expressions for the SU(6) unitary scalar factors and Racah
coefficients related to the Clebsch-Gordan coefficients relevant
for this state. Second, we use these results to estimate masses
for pentaquark states. We indicate how flavour symmetry breaking
may be incorporated in the analysis.

In the next section we present the SU(6)
unitary scalar factors and Racah coefficients, which have been
derived {\em ab initio}. In
section III we recall the nature of the colour-spin hyperfine
interaction while in the following section we use it to estimate
the hyperfine energies for baryons, mesons, diquarks, and
triquarks. In section V the different threads are brought together for
estimating pentaquark masses. In section VI we discuss the results.
We end in section VII with our conclusions.

\section{Some group-theoretic results}\label{s:grp} 
In this section, we collect some results about SU(6) unitary scalar
factors and Racah-like coefficients which will be useful for the
subsequent discussion. Though our motivation in obtaining these
results is the triquark state, they may find some use in
other applications of the SU(6) group.

\subsection{SU(6) unitary scalar factors}
To minimise the complexities, we first
summarize the notations.  A member of an SU(2) multiplet is
denoted by $\{(2I+1),I_3\}$; {\em e.g.,} the  $s_z = +{1 \over
2}$ state of a spin-half particle is $\{{2}, +{1 \over 2}\}$.

For SU(3), the sub-representations are designated by the
SU(2)$^c$ representation\footnote{The superscript `$c$' has been
added to indicate the subgroups of SU(3).} and the `hypercharge',
$Y^c$. Thus, one uses the combination $\{R_3,\alpha, I_3^c\}$
where $R_3$ is the SU(3) representation and $\alpha \equiv [(2I^c
+ 1),Y^c]$. For illustration, a quark state with $I_3^c = +{1
\over 2}$ and $Y^c = {1 \over 3}$ will be denoted as  
$\{3,[{2}, {1 \over 3}],+{1 \over 2}\}$.  

Putting the above together,
an SU(6) state is denoted by
$(R_6,\{R_3,\alpha, I_3^c\},\{(2I+1),I_3\})$ where $R_6$ is the
SU(6) representation while $\{R_3,\alpha,I_3^c\}$ and
$\{(2I+1),I_3\}$ characterize the corresponding SU(3) and SU(2)
sub-representations. The quark state mentioned above, will be
$(6,\{3,[{2}, {1 \over 3}], +{1 \over 2}\}, \{{2},\pm{1 \over
2}\})$, where the SU(3) (SU(2)) quantum numbers are enclosed in
the first (second) braces. In most of the following, it will be
possible to suppress $\alpha, I_3^c$ and $I_3$ -- {\em e.g.,} the
quark state $\equiv$ (6,3,2). This is because the unitary scalar
factors and the Racah coefficients are independent of $\alpha,
I_3^c$ and $I_3$.

The SU(6) unitary scalar factors are generalisations of the
SU(3) isoscalar factors.  The Clebsch-Gordan (CG)  coefficients
of SU(2) are well known. If $ i
\otimes  j =  k \oplus
\ldots $, where  $i,j,k$ are SU(2) representations,  we use
$CG(SU(2)_{i,j,k})$ as an abbreviation for the usual
$C^{i,j,k}_{i_3,j_3,k_3}$ \cite{co}. 
 
Using the SU(2) submultiplets within an SU(3) representation, the
CG coefficients for SU(3) can be expressed in
terms of products of isoscalar factors and SU(2) CG coefficients.
Schematically, for the case $ P
\otimes  Q =  R \oplus
\ldots $:
\begin{equation}
CG(SU(3)_{P,Q,R}) = \left[\matrix
{P & Q & R  \cr
\alpha_P & \alpha_Q & \alpha_R}
\right] \times CG(SU(2)_{I_P, I_Q, I_R }),
\end{equation}

where the $\alpha_i, ~i=P,Q,R$ indicate the sub-representations of
the SU(3) representations $P,Q,R$. The first factor on the
right-hand-side is the SU(3) isoscalar factor. It is independent
of $I_{P3},I_{Q3},I_{R3}$. Tables of SU(3) isoscalar factors 
have been available for long \cite{dS}.

Similarly, in SU(6), if $ X \otimes  Y =  Z \oplus
\ldots $ then 
\begin{eqnarray}
CG(SU(6)_{X,Y,Z}) &=& \left[\matrix
{X & Y & Z\cr
(P_X,I_X) & (P_Y,I_Y) & (P_Z,I_Z)}
\right] \nonumber \\ 
& & \nonumber \\
& & \times CG(SU(3)_{P_X,P_Y,P_Z}) \times
CG(SU(2)_{I_X,I_Y,I_Z}). 
\end{eqnarray}
Here, the first factor on the right-hand-side is an SU(6) unitary
scalar factor -- the generalization of the SU(3) isoscalar
factor.  $P_X (I_X)$ indicates the SU(3) (SU(2)) sub-representation
within the SU(6) multiplet $X$.

Since the triquark state is made out of two quarks ($q_1, q_2$)
and an antiquark ($\bar{q}_3$), the following SU(6) combinations
arise:
\begin{equation}
qq {\rm ~ state:~~} 6 \otimes  6 =  21 \oplus  15 
\label{qq}
\end{equation}
\begin{equation}
qq\bar{q} {\rm ~ state:~~} 21 \otimes  \bar{6} =  120 \oplus  6_1^\phi, \;\; 
15 \otimes  \bar{6} =  84 \oplus  6_2^\phi.
\label{qqbarq}
\end{equation}
or, alternatively,
\begin{equation}
q\bar{q} {\rm ~ state:~~} 6 \otimes  \bar{6} =  35 \oplus  1 
\label{qbarq}
\end{equation}
\begin{equation}
q\bar{q}q {\rm ~ state:~~} 35 \otimes  6 =  120 \oplus  84 \oplus
6_1^\psi, \;\; 1 \otimes  6 =   6_2^\psi.
\label{qbarqq}
\end{equation}
The superscripts $\phi$ and $\psi$ will be clarified in the next
subsection where we identify the Racah coefficients which relate
$(6_1^\phi, 6_2^\phi)$ to $(6_1^\psi, 6_2^\psi)$.
 
For the purpose of the triquark, the SU(6) CG coefficients 
for the product  21 $\otimes$  $\bar{6}$
=  120 $\oplus$  6 are necessary. We have not been able to find the
SU(6) unitary scalar factors for this product in the published
literature \cite{cm}.  Here, therefore, their  {\em ab initio}
calculated values are presented. We follow the generalized
Condon-Shortley phase convention \cite{bb} and obtain:
\begin{equation}
\left[\matrix
{21 & \overline{6} & 6 \cr
(6,3) & (\overline{3},2) & (3,2)}
\right] = \sqrt{6 \over 7},\;\;\;
\left[\matrix
{21 & \overline{6} & 6\cr
(\overline{3},1) & (\overline{3},2) & (3,2)}
\right] = \sqrt{1 \over 7}.
\label{usf1}
\end{equation}

Also,
\begin{equation}
\left[\matrix
{21 & \overline{6} & 120 \cr
(6,3) & (\overline{3},2) & (3,2)}
\right] = \sqrt{1 \over 7},\;\;\;
\left[\matrix
{21 & \overline{6} & 120\cr
(\overline{3},1) & (\overline{3},2) & (3,2)}
\right] = -\sqrt{6 \over 7}.
\label{usf2}
\end{equation}

For the sake of completeness, the SU(6) unitary scalar
factors for the case  15 $\otimes$  $\bar{6}$
=  84 $\oplus$  6 are:
\begin{equation}
\left[\matrix
{15 & \overline{6} & 6\cr
(6,1) & (\overline{3},2) & (3,2)} 
\right] = \sqrt{2 \over 5},\;\;\;
\left[\matrix
{15 & \overline{6} & 6\cr
(\overline{3},3) & (\overline{3},2) & (3,2)}
\right] = \sqrt{3 \over 5}.
\label{usf3}
\end{equation}
and
\begin{equation}
\left[\matrix
{15 & \overline{6} & 84\cr
(6,1) & (\overline{3},2) & (3,2)} 
\right] = \sqrt{3 \over 5},\;\;\;
\left[\matrix
{15 & \overline{6} & 84\cr
(\overline{3},3) & (\overline{3},2) & (3,2)}
\right] = -\sqrt{2 \over 5}.
\label{usf4}
\end{equation}
\vskip 30pt

\subsection{Racah coefficients for the triquark cluster}
\subsubsection{SU(2) and SU(3)} 

In this subsection, after recapitulating the concept of Racah
coefficients, 
using angular momentum as an illustration, the necessary results
useful for the triquark case are presented.

When three angular momenta $j_1,j_2,j_3$ are added, one can
obtain the same final angular momentum $j$ by, for example, (a)
combining $j_1$ and $j_2$ first to get $j_{12}$ and adding $j_3$
to it, or by (b) first adding $j_1$ and $j_3$ to obtain $j_{13}$
and then combining it with $j_2$, or by (c) adding $j_2$ and
$j_3$ to obtain $j_{23}$ and then adding $j_1$ to it.  The states
of the representation $j$ obtained by these three different
routes, may be denoted by $|j_1,j_2,j_3;j_{12},j,m\rangle$,
$|j_1,j_2,j_3;j_{13},j,m\rangle$, and
$|j_1,j_2,j_3;j_{23},j,m\rangle$, respectively. These three
sets of states are related to each other by unitary
transformations whose coefficients, $U$, are called the {\em
normalized} Racah coefficients. For example,

\begin{equation}
U(j_1,j_2,j_3,j;j_{12},j_{13}) =
\langle j_1,j_2,j_3;j_{12},j,m|j_1,j_2,j_3;j_{13},j,m\rangle  .
\label{eq:rdef}
\end{equation}

The triquark state is of the structure ($q_1 q_2 \bar{q}_3$). 
Since the quarks (antiquarks) transform as 6 ($\bar{6}$) of
colour-spin SU(6), for the analysis of these states one requires
the Racah coefficients for SU(6) for the product $6
\times 6 \times  \bar{6}$. 

For most purposes,
it actually suffices if one has the colour SU(3) and spin SU(2)
Racah coefficients. 

The same final triquark state may be reached
by first combining $q_1$ and $q_2$ (colour: $3 \times 3 = \bar3 +
6$ and spin: $2 \times 2 = 3 + 1$) and then combining with each
of these possibilities the antiquark state $\bar{q}_3$. An
alternate way of obtaining the same state is to first pair $q_1$
with $\bar{q}_3$ (colour: $3 \times \bar3 = 8 + 1$ and spin: $2
\times 2 = 3 + 1$) and then adjoining $q_2$ to the result. A
third possibility is obtained by interchanging $q_1
\leftrightarrow q_2$ in the previous alternative. 

We concentrate, in the interest of the pentaquark application, on
the triqark state which transforms like a colour SU(3) triplet
and an SU(2) doublet. The basis states in this sector may be
denoted as:
\begin{equation}
\left( \matrix{|\phi_1\rangle\cr |\phi_2\rangle\cr
|\phi_3\rangle\cr |\phi_4\rangle\cr }
\right) \equiv 
\left( \matrix{|(q_{1}q_{2})^{\bar{3}}_{1}(\bar q_{3})^{\bar 3}_{2}
\rangle_{(3,2)} 
\cr |(q_{1}q_{2})^{6}_{1}(\bar q_{3})^{\bar 3}_{2} \rangle_{(3,2)} \cr
|(q_{1}q_{2})^{\bar{3}}_{3}(\bar q_{3})^{\bar 3}_{2} \rangle_{(3,2)}
\cr |(q_{1}q_{2})^{6}_{3}(\bar q_{3})^{\bar 3}_{2} \rangle_{(3,2)} \cr}
\right)
\label{base1}
\end{equation}
and
\begin{equation}
\left( \matrix{|\psi_1\rangle\cr |\psi_2\rangle\cr
|\psi_3\rangle\cr |\psi_4\rangle\cr }
\right) \equiv
\left( \matrix{|(q_{1}\bar{q}_{3})^{1}_{1}(q_{2})^{3}_{2} \rangle_{(3,2)} 
\cr |(q_{1}\bar{q}_{3})^{8}_{1}(q_{2})^{3}_{2} \rangle_{(3,2)} \cr
|(q_{1}\bar{q}_{3})^{1}_{3}(q_{2})^{3}_{2} \rangle_{(3,2)} 
\cr |(q_{1}\bar{q}_{3})^{8}_{3}( q_{2})^{3}_{2} \rangle_{(3,2)} \cr}
\right),\;\;\; 
\left( \matrix{|\chi_1\rangle\cr |\chi_2\rangle\cr
|\chi_3\rangle\cr |\chi_4\rangle\cr }
\right) \equiv
\left( \matrix{|(q_{2}\bar{q}_{3})^{1}_{1}(q_{1})^{3}_{2} \rangle_{(3,2)} 
\cr |(q_{2}\bar{q}_{3})^{8}_{1}(q_{1})^{3}_{2} \rangle_{(3,2)} \cr
|(q_{2}\bar{q}_{3})^{1}_{3}(q_{1})^{3}_{2} \rangle_{(3,2)} 
\cr |(q_{2}\bar{q}_{3})^{8}_{3}( q_{1})^{3}_{2} \rangle_{(3,2)} \cr}
\right).
\label{base2}
\end{equation}
The notation used here, for example, is that the triquark state with
SU(3) (SU(2)) multiplicity $c'$ ($s'$) obtained through the diquark
combination $(q_{1}q_{2})$ with SU(3) and SU(2) multiplicity $c$ and
$s$, respectively, is represented as $|(q_{1}q_{2})^{c}_{s} (\bar
q_{3})^{\bar 3}_{2}\rangle_{(c',s')}$.

These
possibilities are related by Racah-like coefficients which are
found by explicit calculation to be:
\begin{equation}
\left( \matrix{|\phi_1\rangle\cr |\phi_2\rangle\cr
|\phi_3\rangle\cr |\phi_4\rangle\cr }
\right) 
=\left( \matrix{-\frac{1}{2\sqrt3} &  \frac{1}{\sqrt6} & \frac{1}{2}
& -\frac{1}{\sqrt2} \cr
\frac{1}{\sqrt6} &  \frac{1}{2\sqrt3} & -\frac{1}{\sqrt2}
& -\frac{1}{2} \cr
\frac{1}{2} &  -\frac{1}{\sqrt2} & \frac{1}{2\sqrt3}
& -\frac{1}{\sqrt6} \cr
-\frac{1}{\sqrt2} &  -\frac{1}{2} & -\frac{1}{\sqrt6}
& -\frac{1}{2\sqrt3} \cr}
\right)
\left( \matrix{|\psi_1\rangle\cr |\psi_2\rangle\cr
|\psi_3\rangle\cr |\psi_4\rangle\cr }
\right) 
\label{conv1}
\end{equation}
and

\vskip 10pt

\begin{equation}
\left( \matrix{|\phi_1\rangle\cr |\phi_2\rangle\cr
|\phi_3\rangle\cr |\phi_4\rangle\cr }
\right)
=\left( \matrix{-\frac{1}{2\sqrt3} &  \frac{1}{\sqrt6} & \frac{1}{2}
& -\frac{1}{\sqrt2} \cr
-\frac{1}{\sqrt6} &  -\frac{1}{2\sqrt3} & \frac{1}{\sqrt2}
& \frac{1}{2} \cr
-\frac{1}{2} &  \frac{1}{\sqrt2} & -\frac{1}{2\sqrt3}
& \frac{1}{\sqrt6} \cr
-\frac{1}{\sqrt2} &  -\frac{1}{2} & -\frac{1}{\sqrt6}
& -\frac{1}{2\sqrt3} \cr}
\right)
\left( \matrix{|\chi_1\rangle\cr |\chi_2\rangle\cr
|\chi_3\rangle\cr |\chi_4\rangle\cr }
\right).
\label{conv2}
\end{equation}

\subsubsection{SU(6) Racah coefficients}

One can use the 
unitary scalar factors in eqs. (\ref{usf1}) - (\ref{usf2}) to write:
\begin{equation}
|q_1q_2\bar{q}_3\rangle_{(6_1^\phi,3,2)} = \sqrt{6 \over 7}
~|\phi_4\rangle +
\sqrt{1 \over 7} ~|\phi_1\rangle,
\;\;|q_1q_2\bar{q}_3\rangle_{(120,3,2)} = \sqrt{1 \over 7}
~|\phi_4\rangle -
\sqrt{6 \over 7} ~|\phi_1\rangle.
\label{tri1}
\end{equation}

From eqs. (\ref{usf3}) - (\ref{usf4}) the states obtained if the
diquarks are in the 15 of SU(6) are:
\begin{equation}
|q_1q_2\bar{q}_3\rangle_{(6_2^\phi,3,2)} = \sqrt{2 \over 5}
~|\phi_2\rangle +
\sqrt{3 \over 5} ~|\phi_3\rangle,
\;\;|q_1q_2\bar{q}_3\rangle_{(84,3,2)} = \sqrt{3 \over 5}
~|\phi_2\rangle - \sqrt{2 \over 5} ~|\phi_3\rangle.
\label{tri2}
\end{equation}

Using eq. (\ref{conv1}) one then has:
\begin{equation}
|q_1q_2\bar{q}_3\rangle_{(6_1^\phi,3,2)} = -\sqrt{7 \over 12}|\psi_1 \rangle
-\sqrt{ 2\over 21}|\psi_2 \rangle -\sqrt{ 1\over 28}|\psi_3 \rangle
-\sqrt{ 2\over 7}|\psi_4 \rangle
\label{eq:six1}
\end{equation}
and
\begin{equation}
|q_1q_2\bar{q}_3\rangle_{(6_2^\phi,3,2)} = \sqrt{5 \over 12}|\psi_1 \rangle
-\sqrt{ 2\over 15}|\psi_2 \rangle -\sqrt{ 1\over 20}|\psi_3 \rangle
-\sqrt{ 2\over 5}|\psi_4 \rangle
\label{eq:six2}
\end{equation}
Thus, one arrives at the Racah coefficients:
\begin{equation}
\left( \matrix{|(6_1^\phi,3,2)\rangle \cr |(6_2^\phi,3,2)\rangle}
\right)
=\left( \matrix{\sqrt{\frac{5}{12}} & -\sqrt{\frac{7}{12}}
\cr
\sqrt{\frac{7}{12}} & \sqrt{\frac{5}{12}} }
\right)
\left( \matrix{|(6_1^\psi,3,2)\rangle \cr |(6_2^\psi,3,2) \rangle}
\right)
\label{racah6}
\end{equation}

The non-trivial unitary scalar factors corresponding to eq.
(\ref{qbarqq}) can be written as:
\begin{equation}
\left[\matrix
{35 & 6 & \alpha \cr
i & ({3},2) & (3,2)}
\right] = U_{i,\alpha},
\end{equation}
with $i = 1,2,3$ corresponding to (8,1), (1,3), and (8,3) while
$\alpha = 1,2,3$ to 120, 84, and $6_1^\psi$. Then,
\begin{equation}
U
=\left( \matrix{
-\sqrt{\frac{9}{28}} &  -\sqrt{\frac{8}{21}}  & \sqrt{\frac{25}{84}}  \cr
\sqrt{\frac{9}{20}} &  -\sqrt{\frac{8}{15}}  & -\sqrt{\frac{1}{60}}  \cr
-\sqrt{\frac{8}{35}} &  -\sqrt{\frac{3}{35}}  & -\sqrt{\frac{24}{35}}  \cr}
\right),
\label{usfpsi}
\end{equation}

Now we turn to the application of these results to the pentaquark.

\section{Colour-spin hyperfine interaction}

Besides colour electric forces between all quarks and antiquarks,
there exists a colour-spin hyperfine (colour magnetic)
interaction \cite{dgg}.  In the KL model, it is assumed
that this interaction is operative inside the clusters but, due
to the larger separation, the hyperfine interaction between
clusters is negligible\footnote{Inclusion of the inter-cluster
hyperfine interaction has also been considered \cite{cheung}.}.
The colour-spin SU(6) hyperfine interaction energy is:
\begin{equation}
V = -\sum_{i > j} v_{ij}
(\vec{\sigma_i}.\vec{\sigma_j})(\vec{\lambda_i} . \vec{\lambda_j}).
\label{eq:hyperf}
\end{equation}
Here, $\vec{\sigma}$ and $\vec{\lambda}$ are the Pauli and Gell-Mann
matrices, and $i$ and $j$ run over the constituent quarks and
antiquarks. The common practice is to take $v_{ij} \equiv v$
(flavour symmetry).  $v$ captures information about the radial
dependence of the bound state wave-function. For a composite
system of $n_q$ quarks and $n_{\bar{q}}$ antiquarks, the
hyperfine energy contribution is given by:
\begin{equation}
E_{hyp} = \left[D(q+\bar{q}) - 2D(q) - 2D(\bar{q}) + 16(n_q +
n_{\bar{q}}) \right] v/2,\label{eq:ehyp}
\end{equation}
where
\begin{equation}
D(R_6,R_3,s) = C_6(R_6) - C_3(R_3) - {8 \over 3}s(s+1).
\end{equation}
$C_6$ and $C_3$ are the quadratic Casimir
operators of SU(6) and SU(3) respectively, and $s$, is the spin
of the state.  The effect of this hyperfine interaction on
multiquark exotic states has been a topic of research over
several decades \cite{jaffe,hs}.

The mass estimate for the pentaquark proceeds along the following
pattern. There are three contributions: (a) the masses of the
constituent quarks, (b) the colour-spin hyperfine energy, and (c)
the energy due to the P-wave excitation. The practice has been to
estimate (a) from the masses of the decay products, (baryon +
meson), since their quark content is the same as that of the
parent; but here the hyperfine interaction contribution to the
baryon and meson mass must be first subtracted out, as detailed
in section V. Thus, the
hyperfine interaction enters directly in (b) and also indirectly
in (a) through the way it is extracted.

\section{Hyperfine energies}
\subsection{Mesons and Baryons}
As noted, the hyperfine interaction contributions to the
meson $(q\bar{q})$ and baryon $(qqq)$ masses are required for the
estimation of the pentaquark mass. These can be readily
calculated using eq.  (\ref{eq:ehyp}). For example, in the
flavour symmetry limit, one finds:
\begin{equation}
E_{N(70,1,2)} = -8v,\;\;E_{\Delta(20,1,4)} = 8v,\;\;E_{\pi(1,1,1)} = 
-16v,\;\;E_{\rho(35,1,3)} =\frac{16}{3}v, 
\label{eq:had}
\end{equation}
where in the parentheses the SU(6), SU(3), and SU(2) properties
of the particle have been indicated. 

\subsection{The diquark cluster}
As already mentioned, the  diquark $(qq)$ is usually chosen to be
in the (21,$\bar3$,1) representation which is symmetric in SU(6).
In addition, a diquark can be in the (21,6,3), (15,6,1), and
(15,$\bar3$,3) but these have higher energy. One finds from eq.
(\ref{eq:ehyp}) that the hyperfine energies for these four states
are:
\begin{equation}
E_{(21,\bar{3},1)} = -8v,\;\; E_{(21,6,3)}= -\frac{4}{3}v, \;\;
E_{(15,6,1)}=4v, \;\; E_{(15,\bar{3},3)}= \frac{8}{3}v.
\label{eq:diq}
\end{equation}

\subsection{The triquark cluster}
The triquark cluster in the Karliner-Lipkin model is a member of
the (6,3,2) multiplet and contains two quarks and an antiquark.
The two quarks are assumed to combine to a symmetric 21 of
colour-spin SU(6). For SU(6) 21 $\otimes \bar6$ = 6 $\oplus$ 120,
and the triquark (120,3,2) carries higher hyperfine energy. If
the two quarks are combined in an antisymmetric fashion,
producing a 15 of SU(6), then\footnote{In SU(6), $15 \otimes
\bar6 = 6 \oplus 84$.  In the absence of flavour symmetry, the
triquark is a superposition of these and the 6 and 120 (see
later).} the triquark can be in (6,3,2) or (84,3,2).

More important is the fact that in the existing literature, the
triquark in the (6,3,2) is {\em assumed} to be made with the two
quarks within the cluster forming a (21,6,3). In actuality, so
long as flavour symmetry of the hyperfine interaction holds, the
lowest energy eigenstate of SU(6) receives contributions from
both the $(21,6,3)$ and the $(21,\bar3,1)$ combinations -- see
eq.  (\ref{usf1}) -- and this triquark has the form given in
the first expression in eq. (\ref{tri1}). The other possible
triquark states are the second expression in eq. (\ref{tri1}) and
the  ones in eq. (\ref{tri2}).

\subsubsection{The triquark hyperfine energy}

The calculation of the triquark hyperfine energy using
eq. (\ref{eq:ehyp}) is complicated by the fact that the operator
$D(q+\bar{q})$ and $D(q)$ do not commute; e.g., in eq. (\ref{tri1})
an eigenstate of $D(q+\bar{q})$ is expressed as a linear combination
of those of $D(q)$. 

To circumvent this difficulty, we use the following procedure. We
consider the contribution of eq. (\ref{eq:hyperf}) for the triquark
state term by term as:
\begin{equation}
V =  V_{12} (\vec{\sigma_1}.\vec{\sigma_2})(\vec{\lambda_1} .
 \vec{\lambda_2}) +
V_{13} (\vec{\sigma_1}.\vec{\sigma_3})(\vec{\lambda_1} .
\vec{\lambda_3}) + V_{23}
(\vec{\sigma_2}.\vec{\sigma_3})(\vec{\lambda_2} .
\vec{\lambda_3}).  
\label{eq:thf}
\end{equation}

The hyperfine energy from each term is most readily
calculated in the basis where the two contributing
quarks/antiquarks are first combined \cite{hs2}; i.e., corresponding to the
three terms in the r.h.s. of eq. (\ref{eq:thf}) these are the
$|\phi \rangle$, $|\psi \rangle$, and $|\chi \rangle$ bases
of Sec. \ref{s:grp}, respectively. They are related to each other
through eqs. (\ref{conv1}) and (\ref{conv2}).
In terms of these basis states, one can immediately write down
the expectation value of the Hamiltonian in eq. (\ref{eq:thf}).
Thus\footnote{This form was noted in  \cite{hs2}}, one has:
\begin{equation}
\langle\phi | V|\phi\rangle
=\left( \matrix{\frac{4}{3}V_{12}+ \frac{20}{3}V^{\phi}_+& 4\sqrt{2}V^{\phi}_- & 
\frac{10}{\sqrt{3}}V^{\phi}_- & 2\sqrt{6}V^{\phi}_+ \cr
4\sqrt{2}V^{\phi}_- & -\frac{8}{3}V_{12}+ \frac{8}{3}V^{\phi}_+ & 2\sqrt{6}V^{\phi}_+ 
&\frac{4}{\sqrt{3}}V^{\phi}_-  \cr
\frac{10}{\sqrt{3}}V^{\phi}_- & 2\sqrt{6}V^{\phi}_+  & -4V_{12} 
& 0 \cr
2\sqrt{6}V^{\phi}_+ &  \frac{4}{\sqrt{3}}V^{\phi}_-  & 0
& 8V_{12} \cr}
\right),
\label{expH1}
\end{equation}
where $V^{\phi}_{\pm} = V_{13} \pm V_{23}$. Analogously,
\begin{equation}
\langle\psi | V|\psi\rangle
=\left( \matrix{\frac{8}{3}V_{12}+ \frac{2}{3}V_{13}+ \frac{28}{3}V_{23}& 
\frac{16}{3\sqrt{2}}V^{\psi}_- & 
\frac{4}{\sqrt{3}}V_{12} - \frac{14}{\sqrt{3}}V_{23} &
\frac{8}{\sqrt{6}}V^{\psi}_+
\cr
\frac{16}{3\sqrt{2}}V^{\psi}_- & -\frac{16}{3}V_{13} &
\frac{8}{\sqrt{6}}V^{\psi}_+ & 0 \cr
\frac{4}{\sqrt{3}}V_{12} - \frac{14}{\sqrt{3}}V_{23} &
\frac{8}{\sqrt{6}}V^{\psi}_+ & -2V_{13} & 0 \cr
\frac{8}{\sqrt{6}}V^{\psi}_+ &  0  & 0 & 16V_{13} \cr}
\right),
\label{expH2}
\end{equation}
where $V^{\psi}_{\pm} = V_{12} \pm V_{23}$. $\langle\chi |
V|\chi\rangle$ is similar and is not presented here.
 
The eigenvalues and eigenvectors of this matrix
give the triquark energy and its corresponding group theoretic
configuration, respectively.

The method which we follow can be smoothly adopted to the case of
flavour symmetry violation by appropriately changing the individual coupling
strengths in the three terms of eq. (\ref{eq:thf}). In the
flavour symmetry limit, $V_{12} = V_{23} = V_{13} = v$, whence
$V_-^{\phi} = V_-^{\psi} = 0$. It is seen from 
eq. (\ref{expH1}) that $(\phi_1, \phi_4)$ decouple from
$(\phi_2, \phi_3)$ in this limit.

\section{Pentaquark masses}
\subsection{Hyperfine interaction couplings}
Needless to say, the strength of the colour-spin hyperfine
interaction, $v$, is an important ingredient of the pentaquark mass
estimation. The procedure has generally been to assume that it
takes a universal value which is estimated by ascribing the
$\Delta - N$ mass splitting to this interaction. Using eq.
(\ref{eq:had}),
\begin{equation}
v_3 = \frac{m_\Delta - m_N}{16} \simeq 18.3\; {\rm MeV}.
\label{eq:vbar}
\end{equation}
While this can be a first approximation, it should be borne in
mind that $v$ is determined by the radial dependence of the bound state
wave-function and thus is  most likely different 
for two-body and three-body bound states. Indeed, using eq.
(\ref{eq:had}) for the meson sector one has, 
\begin{equation}
v_2 = \frac{m_\rho - m_\pi}{64/3} \simeq 29.6\; {\rm MeV}.
\label{eq:vmes}
\end{equation}
This is actually an overestimate of $v_2$ since it is well known
that  the pion mass is too small for a simple quark model
interpretation. Eq.  (\ref{eq:vmes}) is only for the purpose of
illustration\footnote{We extract $v_2$ from heavier mesons in the
next subsection.}.  However, it does indicate that it may not be
unreasonable to expect that  $v_2 \neq v_3$ would give a better
approximation to reality.  In the following, in addition to
discussing the results for the choice $v_2 = v_3$,  for the
sake of comparison, we also use a $v_2$ for the diquarks
different from the $v_3$ for the triquarks.

\subsection{Flavour symmetry breaking}
In the limit of exact flavour symmetry, the splitting between the
lowest lying pseudoscalar mesons and the corresponding vector
mesons with the same quark content would be flavour independent.
A measure of flavour symmetry breaking can be obtained from
\begin{equation}
x_f = \frac{m_{K^*} - m_K}{m_\rho - m_\pi} \simeq 0.63.
\label{eq:flav}
\end{equation}
This suggests that the hyperfine interaction involving an
$s$-quark or antiquark carries a suppression by the
factor $x_f$. 
In eqs. (\ref{eq:vmes}) and (\ref{eq:flav}) the use of $m_\pi$
makes the precise values inaccurate. To improve upon this, we use
the masses of the heavier mesons $\rho$, $\phi$, $K^*$, and $K$.
Using eq. (\ref{eq:had}), the hyperfine contributions for these
states are, respectively,  
\begin{equation}
E_{\rho} =\frac{16}{3}v_2, \;\;E_{\phi}=\frac{16}{3}x_{f2}^2v_2,
\;\; E_{K^*} =\frac{16}{3}x_{f2}v_2, \;\;E_{K} =-16x_{f2}v_2,
\label{eq:mesf}
\end{equation}
Here we have added a subscript to $v$ and $x_f$ to indicate that
these values of the hyperfine parameters apply for
two-quark and/or antiquark systems. Using the masses of the mesons, one
can solve for the hyperfine interaction parameters $(v_2,
x_{f2})$ as well as the quark masses. In this manner, one gets:
\begin{equation}
v_2 = 23.62 ~{\rm MeV}, \;\; x_{f2} = 0.782, \;\; 
m_{u,d} = 322 ~{\rm MeV}, \;\; m_s = 471 ~{\rm MeV}.
\label{eq:mespar}
\end{equation}
These values are used in our subsequent calculations.

There are two three-body systems which enter in this analysis.
One is the triquark state and the other the baryon to which the
pentaquark decays. Just as for mesons, one can estimate the
values of $v_3$ and $x_{f3}$ from the $N - \Delta$ and $\Sigma -
\Sigma^*$ mass splittings which are given by:
\begin{equation}
E_\Delta - E_N = 16v_3,\;\; E_{{\Sigma}^*} - E_{\Sigma} =
\frac{16}{3}v_3(2x_{f3} + 1),\;\; E_{{\Xi}^*} - E_{\Xi} =
\frac{16}{3}v_3x_{f3}(x_{f3} + 2).
\label{eq:barf}
\end{equation}
As a consistency check, we use the
values so obtained to calculate the $\Xi - \Xi^*$ splitting and
find that the agreement is not satisfactory. Therefore, we use all
of the three above splittings to arrive at the best-fit values:  
\begin{equation}
v_3 = 17.89 ~{\rm MeV}, \;\; x_{f3} = 0.708.
\label{eq:barpar}
\end{equation}
In the following, these have been used for the triquark and
baryons.

\subsection{P-wave excitation}
The energy due to the P-wave excitation can be estimated from the
recently observed $D_s^*$ state at 2317 MeV, which is believed to
be an orbital excitation of the state at 2112 MeV. This
gives\footnote{Alternatively, one might use $ E_{P} =
m_{\Lambda({\frac{1}{2}})^-} - m_{\Lambda({\frac{1}{2}})^+}
\simeq (1406 - 1116)\; {\rm MeV} = 290\; {\rm MeV}$. This will
increase all pentaquark mass estimates below by $\sim $ 85 MeV.}
\begin{equation}
E_{P}= m_{D_s^*}(P) -
m_{D_s^*}(S) \simeq (2317 - 2112)\; {\rm MeV} = 205\; {\rm MeV}.
\label{eq:pw}
\end{equation}
\section{Results}
\subsection{The flavour antidecuplet and the octet}
Putting together the inputs from the previous sections, one can
readily obtain the masses of the pentaquark states in the
Karliner-Lipkin model. For example, for $\Theta^+$, using eqs.
(\ref{eq:had}) and (\ref{eq:diq}):
\begin{equation}
m_{\Theta^+} = \{(m_N + 8v_3) + (m_s + m_q)\} + E_P  - 8v_2 + 
E_{tri}(v_3,x_{f3}),
\label{klmass}
\end{equation}
where the expression in the curly brackets is the contribution
from the quark masses.  The last (penultimate) term is the hyperfine
energy of the triquark (diquark). For other
pentaquarks, the r.h.s. in eq.  (\ref{klmass}) has to be
appropriately modified to reflect the quark content of the state
and, when necessary, deviations from flavour symmetry have to be
incorporated in eq.  (\ref{eq:thf}) to obtain the correct
$E_{tri}(v_3,x_{f3})$.

\begin{table}[hbt]
\begin{center}
\begin{tabular}{|c|c|c|c|c|c|c|c|}
\hline
Pentaquark &\multicolumn{7}{|c|}{Mass (in MeV)} \\ \cline{2-8} 
states & $\Theta^+$ &  $N_{10}$ & $\Sigma_{10}$ & $\Xi_{10}$ &
 $N_{8}$ & $\Sigma_{8}$ &  $\Xi_{8}$  \\ \hline
Lowest    &1601&1358&1626 & 1783&2057& 2217 & 2326\\ \hline 
SU(6) Excited   &1789&1573&1840 & 1966&2321& 2439 & 2512\\ \hline 
\end{tabular}
\caption{\sf \small Pentaquark lowest and first colour-spin
excited state masses for the reference values of the parameters
in eqs. (\ref{eq:mespar}) and (\ref{eq:barpar}).}
\label{t1}
\end{center}
\end{table}

As noted earlier, the pentaquark states fill an octet and an
antidecuplet of flavour. Excepting for the three states, $\Theta^+
\equiv udud\bar{s}$, $\Xi^{--} \equiv dsds\bar{u}$, and $\Xi^{+}
\equiv usus\bar{d}$, all other states in the antidecuplet have
partners in the octet with identical isospin and hypercharge. In
estimating the masses,  we have assumed {\em ideal} mixing between
the partners and ascribed the {\em lighter} member to the
antidecuplet. Note that isospin symmetry is assumed unbroken, so it is
enough to present the mass of one member of an isomultiplet. The
masses of the pentaquark states at the reference values of the
parameters -- see eqs. (\ref{eq:mespar}) and (\ref{eq:barpar}) --
are given in Table 1.

\begin{center}
\begin{figure}[thb]
\includegraphics[width=0.5\textwidth,height=0.4\textheight,angle=270]
{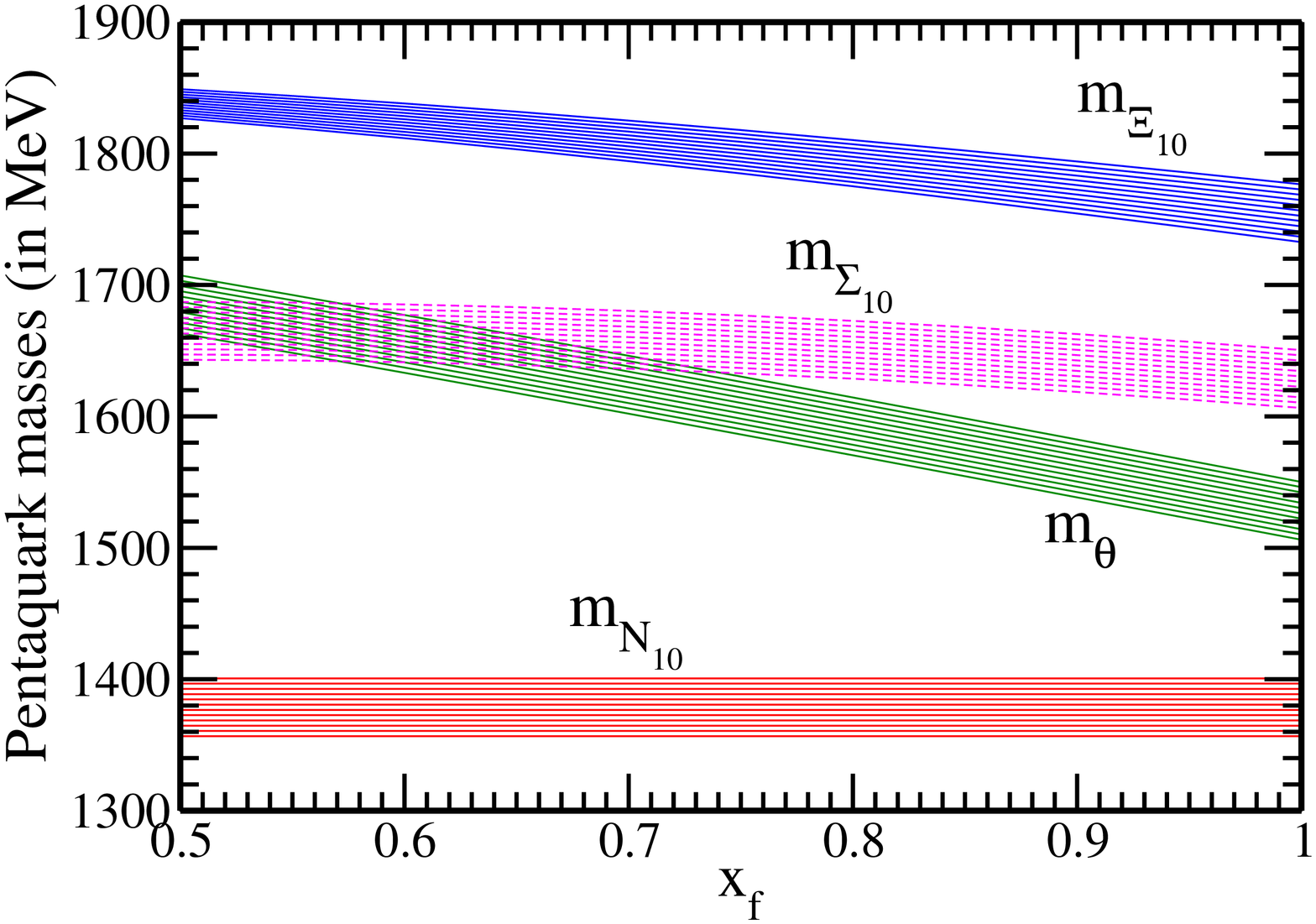}
\vskip -7.8cm
\hskip 8.0cm
\includegraphics[width=0.5\textwidth,height=0.4\textheight,angle=270]
{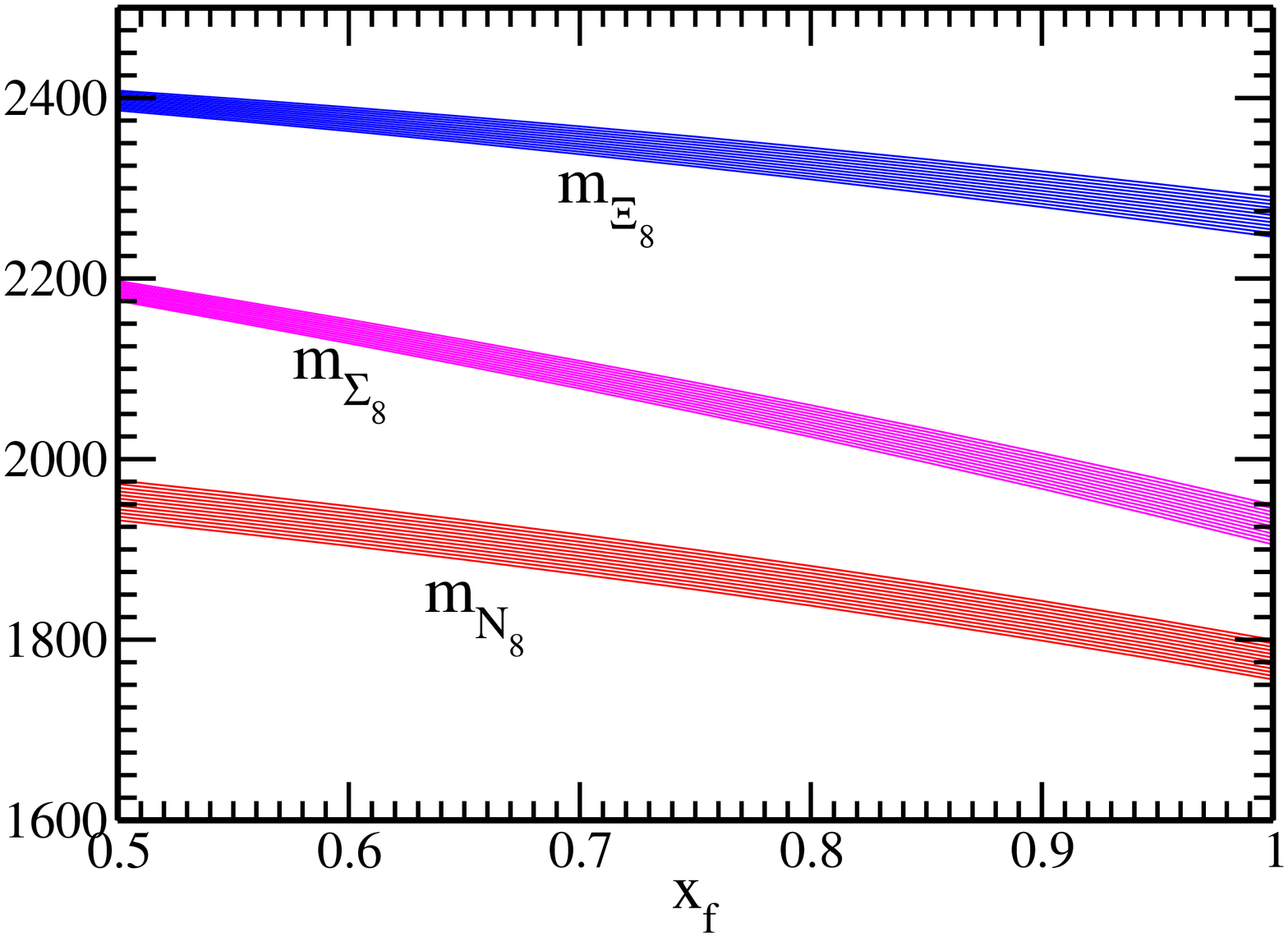}
\caption{\em The dependence of pentaquark masses on the deviation
from flavour symmetry ($x_f = 1$). The left (right) panel
corresponds to flavour antidecuplet (octet) pentaquarks. The
bands are obtained when the diquark hyperfine
interaction strength is varied over the range  17.89 MeV
$\leq v_2 \leq 23.62$ MeV (see text).}
\label{f1}
\end{figure}
\end{center}

In Fig. \ref{f1}, in the left panel the antidecuplet pentaquark
masses are shown as a function of the flavour symmetry violation
parameter $x_f$, which assumes the value unity in the symmetry
limit. In view of the closeness of the estimates of $x_f$ in eqs.
(\ref{eq:mespar}) and (\ref{eq:barpar}), for this figure we have
taken $x_{f3} = x_{f2} = x_{f}$. The triquark interaction
strength has been kept fixed at  $v_3 = 17.89$ MeV. The bands
arise from a variation of the strength of the diquark hyperfine
interaction, $v_2$, with the lower edge corresponding to $v_2 =
v_3$ and the upper to  $v_2 = 23.62$ MeV (see eq.
(\ref{eq:mespar})).  For this figure, $E_P$ has been chosen as 209
MeV, following eq. (\ref{eq:pw}). It is observed that the
triquark corresponding to the lowest eigenvalue of the hyperfine energy
Hamiltonian -- eq. (\ref{expH1}) --  is predominantly a combination
of the states $\phi_1$ and $\phi_4$ (see eq. (\ref{base1})) which
are antisymmetric in the quark flavours.

%%%%%%%%%%%%%%%%%%%%%%%%%%%%%%%%%%%%%%%%%%%%%%%%%%%%%%%%%%%%%%%%%%%
%%%%%%%%%%%%%%%%%%%%%%%%%%%%%%%%%%%%%%%%%%%%%%%%%%%%%%%%%%%%%%%%%%%

\vskip -40pt
Note that, $N_{10}$, the non-strange member of the
antidecuplet\footnote{This state could have been proposed as a
possible interpretation of the Roper resonance at 1440 MeV.} is
predicted to be at a mass of 1355 MeV for $v_2 = v_3$ which is
enhanced to $\sim$ 1400 MeV when $v_2 = 23.62$ MeV is used. This
prediction is independent of the choice of $x_f$ since the state
does not have strange quarks.  For the exotic $\Xi^{--}_{10}$
state the mass prediction is in the range 1795 -- 1825 MeV for
$x_f$ = 0.7 to be compared with that of the experimentally
observed state at 1862 MeV \cite{na49}.

In the right panel of  Fig. \ref{f1} are shown the octet
pentaquark masses.  The splitting between the masses of the octet
states and the corresponding antidecuplet states is seen to be
typically around 500-600 MeV. As noted earlier, at the level of
these calculations, the masses of the I=1 and I=0 members of the
octet with S = -1 are the same.   The non-strange neutral state
in the octet, $N_8^0$, has the quark structure $(ud\bar{s})(ds)$
and its mass is consequently dependent on $x_f$.

A remark needs to be made about the symmetry property of the
triquark state for the octet pentaquarks. This feature is most
easily brought out from a consideration of the S = -2 member of
the octet, $\Xi_8$, which has the quark structure
$(us)(ss\bar{s})$. The diquark is antisymmetric in flavour so its
choice is fixed. Unlike all the other states, here the triquark
is compelled to have two identical ($s$) quarks, besides the
antiquark.  Consequently, in the notation of section II, it can
arise only from a combination of the states $\phi_2$ and $\phi_3$
(see eq. (\ref{base1})) which are symmetric in flavour.
Obviously, all states in the pentaquark octet will share this
feature in the exact flavour SU(3) limit.

The H1 experiment at HERA found evidence of a
possible charmed pentaquark at mass 3099 MeV \cite{hera}. This
state has the quantum numbers of a pentaquark with the structure
$udud\bar{c}$.  Including flavour violation 
($x_f = 0.23$ for the
$\bar{c}$ quark) and taking 
$v_2$ = 23.62 MeV, $v_3$ = 17.89 MeV, we find the
predicted mass for such a state is 2757 MeV.

\subsection{Triquark SU(6) excitations}

Colour triplet, spin $\frac{1}{2}$ triquarks come in four
varieties. These are the four eigenstates of the hyperfine energy
matrix in eq. (\ref{expH1}). The results presented so far are
obtained using the eigenstate with the minimum energy consistent
with symmetry requirements -- a
certain choice of colour-spin assignments for the quark clusters
-- and leads to the lowest lying pentaquarks. It is evident that
the other triquark eigenstate clusters also lead to colour
singlet spin $1 \over 2$ pentaquark states, albeit heavier. How
different are the masses in these other cases?

For illustration, we show in Table \ref{t1} the masses of the
first excited partners of the antidecuplet and octet pentaquarks
for the reference values of the hyperfine interaction parameters.
In the flavour symmetry limit ($x_{f3} = x_{f2}$ = 1), the
spacing between the excited states is independent of the flavour
and the lowest and first excited states are separated by 215 MeV
(370 MeV) for every member of the antidecuplet (octet).
 
There is no obvious argument to suppress the production of these
additional states.  It will be of interest to extend the ongoing
searches to look for such SU(6) colour-spin excited partners, a
novelty of QCD and the pentaquark system.

\section{Conclusions}

A pentaquark interpretation of the $\Theta^+$ leads to
predictions of several other colour singlet states in a similar
mass range which populate an antidecuplet and an octet of flavour
SU(3). In this paper, the masses of these pentaquark states have
been calculated in a triquark-diquark (Karliner-Lipkin) model
with refined estimates, upto first order, of the colour-spin
SU(6) hyperfine interaction contributions.

Motivated by the structure of these states,  the SU(6) unitary
scalar factors relevant for the $qq\bar{q}$ triquark structure
and the Racah coefficients, not available in the literature, have
been calculated {\em ab initio}. Using these results, the
colour-spin SU(6) hyperfine contributions have been obtained
taking two variations from the simplest picture. One of these
concerns the deviation from flavour symmetry. The other
originates from a possible difference in the strength of the
hyperfine interaction for two- and three-quark bound states which
can be related to  the known splittings in baryonic and mesonic
systems. Both of these variations do affect the pentaquark mass
predictions.  An element of uncertainty is introduced in these
mass estimates by the P-wave excitation energy for which we have
used the information from the $D$-meson system.

The triquark states within the antidecuplet and the octet are
chosen, for good reason,  to be the lowest eigenstate of the
hyperfine energy Hamiltonian satisfying  symmetry
requirements.  The other eigenstates are possible triquark states
of SU(6) colour-spin excitations. The masses of colour singlet,
spin $1\over 2$ pentaquarks resulting from these triquark
excitations have also been estimated.

Irrespective of whether the claimed observation of the $\Theta^+$
baryon is vindicated or not, pentaquarks can prove to be the
tip of a revealing iceberg of new hadronic states illuminating
novel facets of QCD.

\vskip 40pt

\parindent 0pt

{\large{\bf {Acknowledgements}}}

\vskip 10pt

In the initial stages, this work was partially supported by the
Department of Science and Technology, India (AR) and the Council
of Scientific and Industrial Research, India (SKM).

\end{document}